\documentclass[twoside,reqno,twocolumn]{article}

\usepackage[letterpaper]{geometry}

\usepackage{ltexpprt}
\usepackage{hyperref}

\usepackage{amsmath,amssymb}  
\usepackage{bm}  
\usepackage{fancyhdr}
\usepackage[at]{easylist}
\usepackage[mathletters]{ucs}
\usepackage[utf8x]{inputenc}
\usepackage{csquotes}

\usepackage{tcolorbox}

\usepackage{cite}

\usepackage{adjustbox}
\usepackage{setspace}
\usepackage{longtable}

\usepackage{algpseudocode}
\usepackage{algorithm}

\usepackage{xpatch}
\makeatletter
\xpatchcmd{\algorithmic}{\itemsep\z@}{\itemsep=1ex plus2pt}{}{}
\makeatother

\title{
Optimal Trajectory Planning 
with Collision Avoidance
for 
Autonomous Vehicle 
Maneuvering
}

\author{Jason Zalev\thanks{Magna International Inc. 
Email: jason.zalev@magna.com}}

\date{\today}

\newcommand{\bfx}{\mathbf x}

\newcommand{\bfz}{\mathbf z}
\newcommand{\bfu}{\mathbf u}

\newcommand{\bfp}{\mathbf p}
\newcommand{\bfq}{\mathbf q}

\newcommand{\bfn}{\mathbf n}

\newcommand{\bfr}{\mathbf r}
\newcommand{\bfb}{\mathbf b}

\newcommand{\scrK}{\mathcal{K}}
\DeclareMathOperator{\p}{p}

\newcommand\norm[1]{\left\lVert#1\right\rVert}

\begin{document}

%
%\newcommand\relatedversion{}
%\renewcommand\relatedversion{\thanks{The full version of the paper can be accessed at \protect\url{https://arxiv.org/abs/TBD}}} % Replace URL with link to full paper or comment out this line

%\setcounter{chapter}{2} % If you are doing your chapter as chapter one,
%\setcounter{section}{3} % comment these two lines out.

%\title{\Large SIAM/ACM Preprint Series Macros for Use With LaTeX\relatedversion}
%\author{Corey Gray\thanks{Society for Industrial and Applied Mathematics.}
%\and Tricia Manning\thanks{Society for Industrial and Applied Mathematics.}}

\date{}

\maketitle

% Copyright Statement
% When submitting your final paper to a SIAM proceedings, it is requested that you include
% the appropriate copyright in the footer of the paper.  The copyright added should be
% consistent with the copyright selected on the copyright form submitted with the paper.
% Please note that "20XX" should be changed to the year of the meeting.

% Default Copyright Statement
%\fancyfoot[R]{\scriptsize{Copyright \textcopyright\ 20XX by SIAM\\
%Unauthorized reproduction of this article is prohibited}}

% Depending on which copyright you agree to when you sign the copyright form, the copyright
% can be changed to one of the following after commenting out the default copyright statement
% above.

%\fancyfoot[R]{\scriptsize{Copyright \textcopyright\ 20XX\\
%Copyright for this paper is retained by authors}}

\fancyfoot[R]{\scriptsize{Copyright \textcopyright\ 2024\\
Copyright retained by principal author's organization}}

\fancyfoot[L]{\scriptsize{SIAM Conference on Control and Its Applications\\
July 28-30th, 2025, Montreal, Canada }}

%\fancyfoot[R]{\scriptsize{Copyright \textcopyright\ 2024}}

%\pagenumbering{arabic}
%\setcounter{page}{1}%Leave this line commented out.

%\IEEEtitleabstractindextext{%
\begin{abstract}

To perform 
autonomous driving
maneuvers, such as parallel or perpendicular parking, 
a vehicle 
requires continual speed and steering adjustments 
to follow a 
%provided
generated
path. 
In consequence, 
the path's quality %can
is a 
limiting factor of %in %of %that limits
the
vehicle maneuver's performance. 
While 
most
path planning 
approaches 
include
finding
a collision-free route, 
optimal
trajectory planning %includes
involves
 solving 
the best 
transition 
from initial to final 
states,
minimizing 
the
 action
over all paths 
permitted
by
a
kinematic model. 
Here
we propose a novel method 
based
 on
sequential convex optimization, 
which 
permits flexible 
and efficient
optimal trajectory generation. 
The objective is to 
achieve 
the
 fastest time, 
shortest distance, 
and
 fewest number of path segments
to satisfy motion requirements, while avoiding sensor blind-spots.  
In our approach, 
vehicle
kinematics are represented by a discretized Dubins 
model. 
To avoid collisions,  
each waypoint is constrained by linear inequalities 
representing 
closest distance 
of
obstacles
to 
a polygon 
specifying 
the vehicle's extent. 
To promote smooth and valid trajectories, 
the solved
kinematic state 
and control 
variables 
are constrained and regularized by 
penalty 
 terms in the model's cost function, 
which enforces
 physical restrictions 
  including limits for steering angle, acceleration and speed. 
In this paper,
we analyze
 trajectories 
obtained 
for
 several parking scenarios.  
Results
demonstrate 
efficient and collision-free
motion
generated by
the proposed technique.

\end{abstract}
%}

%\begin{document}
\maketitle
%\IEEEdisplaynontitleabstractindextext

\section{Introduction}

In autonomous parking systems, 
vehicle
sensors capture continuous information about 
%surroundings, 
%the 
surrounding
obstacles, 
%surrounding 
%that surround
%a vehicle, 
including the presence of 
%parked 
%cars, 
%vehicles, 
%other vehicles,
pedestrians, 
 barriers,
curbs, 
%other vehicles
%and 
lane markings
and other vehicles.  
% and other obstacles. 
%The %position and 
%Proximity 
During parking, 
the proximity 
to these obstacles is used 
%for planning 
to plan
a collision-free 
%path 
route
%to
%that will
%to safely enter or exit 
%for safely entering or exiting 
for entering or exiting 
%for safe entry or exit of
a parking slot. 
%Slot geometry 
Since the geometry of a slot
can vary from one situation to the next, 
%so 
flexible %path-planning %routines are good. 
planning
approaches are needed to ensure optimal functionality in all circumstances.

Maneuvering 
approaches
for automotive systems 
can be categorized into: 
%by: 
direct methods\cite{li2016time, qiu2021hierarchical,chai2018two,li2021optimization,li2019tractor, sun2021successive,zhu2015convex,boyali2020applications, noreen2016optimal,karaman2010optimal,
zips2013fast, 
verschueren2014towards, liu2017path}, 
where the path and trajectory are found simultaneously; and  
indirect methods\cite{gomez2008continuous,  reeds1990optimal, 
lipp2014minimum,
%verschueren2014towards, li2016time, liu2017path,
zhang2020optimization}, 
where maneuvers 
%which
are split into
stages of 
path planning and path following. 
% stages of path planning and path following.  
Indirect methods often use heuristics
with efficient computation to produce a feasible path, 
but the resulting trajectory 
may be sub-optimal. 
This includes
geometric  planners, which involve
creating simple paths using 
spline curves\cite{gomez2008continuous} or 
circular arcs and segments\cite{reeds1990optimal}. 
Direct approaches may
involve higher computation to solve optimal trajectories\cite{li2016time, qiu2021hierarchical,chai2018two,li2021optimization,li2019tractor, sun2021successive,zhu2015convex,boyali2020applications, noreen2016optimal,karaman2010optimal} 
or use fast approximations\cite{zips2013fast}. 
Some approaches involve sequential convex optimization 
\cite{malyuta2022convex,
%%%qiu2021hierarchical,
mao2018successive,
reynolds2020real, 
malyutadiscretization, li2016time, qiu2021hierarchical,chai2018two,li2021optimization,li2019tractor, sun2021successive,zhu2015convex,boyali2020applications
%, verschueren2014towards, liu2017path
}
to handle non-linear constraints and objectives. 
For real-time applications, 
model predictive control (MPC)\cite{verschueren2014towards, liu2017path}
involves performing dynamic updates to path and trajectory, optimizing over a short time horizon for efficiency.

\begin{figure}[b!]
  \centering
  \includegraphics[width=0.8\columnwidth]{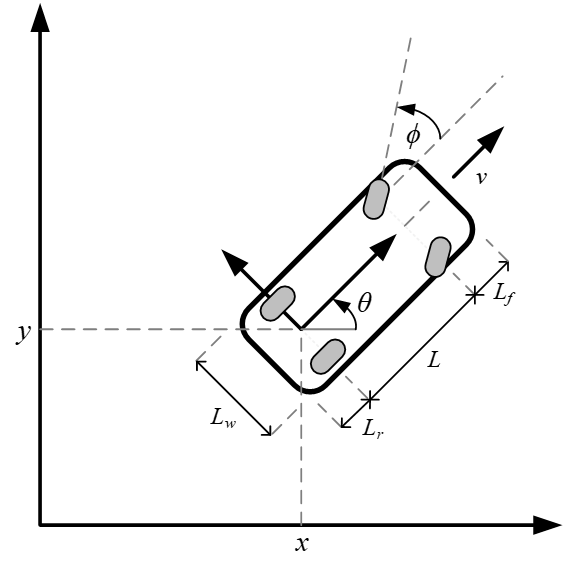}
  \caption{Vehicle geometry}
  \label{fig:fig1}
\end{figure}

In this paper, we propose a
method for optimal
 trajectory planning 
and collision avoidance 
in autonomous vehicles.  
Obstacles are represented by rectangular regions. 
A trajectory is defined by 
a series of 
waypoints 
that specify
the host vehicle's
kinematic state. 
The host vehicle has a
rectangular shape, which  
is extended %to form 
into 
a 
convex
polygon 
to fill 
gaps 
due to 
 the trajectory's curvature. 
Constraints to avoid collision are based on finding the closest distance of 
the polygon to 
each rectangular obstacle, 
 and penalizing waypoints that would cause any overlap. 
Sequential convex optimization minimizes the trajectory's total elapsed-time. 
The kinematic state variables are limited by upper and lower bounds, 
which provides flexibility to control the resulting trajectory's characteristics. 
Additional smoothness constraints 
are used to 
regularize the trajectory, contributing to faster convergence.  
As well, 
our approach proposes
special constraints to 
avoid blind-spots, 
ensuring
that 
low-visibility regions are swept by sensors 
along
the trajectory, 
avoiding
potential collisions.

%\subsection{Contributions}
%
\section{Kinematic Model}

\label{sec:kinematic_model}
\lfoot{}
We model the vehicle's trajectory 
using 
kinematic state 
vector
 $\bfx(t)$ and control input 
 vector
  $\bfu(t)$, 
 % defined as
according to
%equal to 
%  which we model as
% as 
\begin{align}
\bfx(t) = 
\begin{bmatrix}
x(t)\\
y(t)\\
v(t)\\
a(t)\\
\theta(t) \\
\phi(t) \\
\end{bmatrix}
%\end{align}
%
%\begin{align}
,\quad
\bfu(t) = 
\begin{bmatrix}
j(t)\\
\omega(t)
\end{bmatrix},
\label{eq:xt_ut}
\end{align}
%with 2D position coordinates $x(t)$ and $y(t)$, 
%velocity $v(t)$, acceleration $a(t)$, vehicle orientation $\theta(t)$, 
%steering angle $\phi(t)$,  steering rate $\omega(t)$ and jerk $j(t)$ . 
where $x(t)$ and $y(t)$ are 2D position coordinates, 
$v(t)$ is velocity, $a(t)$ is acceleration, $\theta(t)$ is orientation, $\phi(t)$ is steering angle,  $\omega(t)$ is steering rate and $j(t)$ is jerk.

%The state space function is 
%We construct a 
A %non-linear 
state space function 
$f(\bfx(t), \bfu(t), \bfp)$
is constructed
by specifying 
%from 
the time-derivative of $\bfx(t)$, 
%which is equal to  
such that
\begin{align}
f(\bfx(t),  \bfu(t), \bfp)
&=
\frac{d \bfx(t)}{d t}
%\\
%%\frac{d }{d t}
%%\begin{bmatrix}
%%x(t)\\
%%y(t)\\
%%v(t)\\
%%a(t)\\
%%\theta(t) \\
%%\phi(t) \\
%%\end{bmatrix}
%&=
=
\begin{bmatrix}
v(t) \cos\theta(t)\\
v(t) \sin\theta(t)\\
a(t)\\
j(t)\\
%\theta(t) \\
v(t) \tan\phi(t)/L \\ 
\omega(t)
\end{bmatrix},  
\label{eq:f_xu}
\end{align}
%This %equation 
%Equation \eqref{eq:f_xu}
%
%This
which 
constrains the dynamics of motion %which
%such that \eqref{eq:f_xu}
and
must be satisfied by any valid trajectory. 
Here, 
the vector $\bfp$ %consists of parameters 
corresponds to parameters 
that may be adjusted in the model. 
% any valid trajectory must satisfy. 
The constant $L$ is the vehicle's wheel-base, as depicted in Figure~\ref{fig:fig1}.

%Since equation \eqref{eq:f_xu} is non-linear, the optimial 
%Equation 
Since equation~\eqref{eq:f_xu} is non-linear, 
%yielding 
it yields 
a non-convex optimization problem that is difficult to solve directly. 
%Moreover,  minimizing $t_\text{f}$, the time to reach the final state, 
%impacts how way-points are discretized, and this must be accounted for.
Therefore, 
to compute the trajectory, we use sequential convex optimization, 
described in Section~\ref{sec:traj_opt}, 
 to approximate
the original non-convex problem with an iterative sequence of convex sub-problems. 
%, until convergence.  
Each sub-problem is obtained by linearizing the 
%discretized 
trajectory with respect to 
a reference solution, % $(\bar \bfx, \bar \bfu, \bar \bfp)$, 
corresponding to the 
previous iteration's output\cite{reynolds2020real,mao2018successive,malyutadiscretization}. %, 
%output from the previous iteration,
% until convergence. 
%From   equation~\eqref{eq:f_xu}, %linearization of the state space equation is given by 

%By linearizing %the state space function of 
The achieve this, 
%following the approach of~\cite{reynolds2020real},
we linearize
equation~\eqref{eq:f_xu}
with respect to $(\bar \bfx, \bar \bfu, \bar \bfp)$, 
%we have
which yields
\begin{subequations}
\begin{align}
\frac{d \bfx(t)}{d t}
&=
f(\bfx(t),  \bfu(t), \bfp) \nonumber
\\
\begin{split}
& \approx
f(\bar \bfx(t), \bar \bfu(t),\bar \bfp) %\nonumber
%+
%\\
%&\qquad
%f\left(
%\begin{bmatrix}
%\bar \bfx(t) \\  
%\bar \bfu(t)\\
%\bar t
%\end{bmatrix}
%\right)
%\\
%&\quad
%\ldots
\\
&\qquad
+A(t) [\bfx(t)-\bar \bfx(t)] %\nonumber
\\
&\qquad
+B(t) [\bfu(t)-\bar \bfu(t)] %\nonumber
\\
&\qquad
+E(t) [\bfp-\bar \bfp]
%\begin{bmatrix}
%A(t)&B(t)&E(t)
%\end{bmatrix}
%\begin{bmatrix}
%\bfx(t)-\bar \bfx(t) \\  
%\bfu(t) - \bar \bfu(t)\\
%t- \bar t
%\end{bmatrix}
%\\ &\qquad
%+f(\bar \bfx(t), \bar \bfu(t),\bar t)
\end{split}
\label{eq:ABE_xup}
\\&=
\begin{bmatrix}
A(t)&B(t)&E(t)
\end{bmatrix}
\begin{bmatrix}
\bfx(t) \\  
\bfu(t)\\
\bfp
\end{bmatrix}
+\bar \bfr(t),
\label{eq:ABE_xup_r}
\end{align}
where 
%$
\begin{align}
\bar \bfr(t) =
%f(\bar \bfx(t), \bar \bfu(t),\bar \bfp)
f(\bar \bfx, \bar \bfu,\bar \bfp)
-A(t)\bar \bfx(t)
-B(t)\bar \bfu(t)
-E(t) \bar \bfp, 
\label{eq:r_bar_t}
\end{align} 
\end{subequations}
and
\begin{subequations}
\begin{align}
A(t) &= \nabla_\bfx f(\bar \bfx(t),\bar \bfu(t),  \bar \bfp)
\\
B(t) &= \nabla_\bfu f(\bar \bfx(t), \bar \bfu(t), \bar \bfp)
\\
E(t) &= \nabla_\bfp f(\bar \bfx(t), \bar \bfu(t), \bar \bfp). 
\end{align}
\label{eq:grad_ABE}
\end{subequations}
%
%In the above equations, 

In equation~\eqref{eq:ABE_xup_r}, 
the reference components 
$\bar \bfx(t)$, $\bar \bfu(t)$ and $\bar \bfp$
of \eqref{eq:ABE_xup} 
%are
have been
 collected %together
%in \eqref{eq:ABE_xup_r} 
%forming  
%into
%to form
into 
an overall reference vector $\bar \bfr(t)$,  
%in equation~\eqref{eq:ABE_xup_r}, 
%which is given  in equation~
%which 
%that
%remains constant at each $t$,
%as 
defined in
%according to 
\eqref{eq:r_bar_t}. 
%This allows 
%This shows that %the linearization in 
%Thus, 
%This  allows 
%As shown, 
%Here, 
As a consequence, 
equation~\eqref{eq:ABE_xup_r}
isolates 
the current trajectory, 
%, 
%corresponding to 
%$(\bfx(t),\,\bfu(t),\,\bfp)$, 
which 
is
%represents an unknown variable
%corresponding to the unknown variables
solved 
%as unknown
in
each iteration. %, 
%corresponding to 
%$\bfx(t)$, $\bfu(t)$ and $\bfp$.  
%$(\bfx(t),\,\bfu(t),\,\bfp)$.  
%which are isolated
%which have been isolated
%for solving in
%which each iteration must solve. 
%which %must be 
%is 
%The matrices
%In this form, 
%Here, 
Since 
%the matrix functions 
$A(t)$, $B(t)$,  
%and 
$E(t)$
and
$\bar \bfr(t)$ 
  %are constant for each time $t$, 
depend only on the reference components,
%and remain constant at each fixed time $t$, 
%which allows
%allowing
this allows
%allowing these to
%so
%these can 
equations~\eqref{eq:r_bar_t}
%, %of equation~\eqref{eq:grad_ABE}, 
%and the vector function 
and 
\eqref{eq:grad_ABE} 
%can 
to 
be
viewed 
as constants 
that are
pre-computed for each iteration. 
%as constants 
%%pre-computed
%computed
% prior to each iteration.  
%and viewed 
%as constant. 
%. 
%This means 
%In this form, 
%equation~
%in 
%\eqref{eq:ABE_xup_r} 
%is a differential equation that depends on 
%the state, control and parameter vectors 
%to depend 
%depends
%only on %the vector functions
%Thus,
%, corresponding to
%for 
%to be solved for 
% the current trajectory. %solution.  

\section{Discretization}

To solve the 
%unknown 
trajectory, 
%it
equation~\eqref{eq:xt_ut}
 is converted to a discretized form, 
%the discretized form is
%in discrete form it is 
%In discretized form, 
%the trajectory is
%A trajectory, in discretized form, is
%trajectories are
 represented by %a set of 
 state variable 
 $\bfx_k=[x_k, y_k, v_k, a_k, \theta_k, \phi_k]^T$ with 
control variable  
 $\bfu_k=[j_k, \omega_k]^T$,  
% state and control variables $\bfx_k$ and $\bfu_k$,  
%at discrete times
%for %indices 
%each
%at
%each index
%timesteps
where
 $k=1\ldots N$. % and %$N$ is the number of way-points. % in the trajectory. 
% there are $N$ way-points. 
%A sample at 
Here, the discrete
%Each 
%time 
index $k$ 
corresponds to 
%represents
%a 
%sample at the 
%a 
%sample at 
the 
continuous time %sample at
variable
%variable
%at
$t = (k-1)\Delta t$, 
% $t = k \frac{t_f}{N}$,  
where 
%$\Delta t=  \frac{t_\text{f}}{N}$.  
$\Delta t=  t_\text{f}/N$.      
%so the
%such that
%so
%Here, %there are
%The
This %represents
corresponds to
%representing 
%The
 $N$ %samples 
 way-points
 with
%have
  uniform 
temporal  
  spacing, where
%are evenly space over an interval
% and  
$t_\text{f}$ is the time required to reach the final state.  
%In 
%To obtain 
%the 
%In
%discretized form, 
%the 
For
discretization, 
$t_\text{f}$ is included 
%as the only variable
in the parameter set $\bfp$, 
%which allows this
allowing it 
 to be solved as 
 an unknown model variable. 
 %one of the model's unknown variables. 
%a variable in the model. 

For convenience, 
%in the discretized model, 
we 
%stack
concatenate 
%the %discretized
%discrete
% samples
 $\bfx_k$ and $\bfu_k$ %, for  $k=1\ldots N$, 
% from 
% each time index $k$ 
 into 
%concatenated 
the stacked
vectors $\tilde \bfx$ and $\tilde \bfu$, according to
\begin{subequations}
\begin{align}
\tilde \bfx &= \text{vec}(\bfx_1, \bfx_2, \ldots, \bfx_N),
\\
\tilde \bfu &= \text{vec}(\bfu_1, \bfx_2,\ldots, \bfu_N),
\\
\tilde \bfp &= t_\text{f}. 
\end{align}
\label{eq:xup_stacked}
\end{subequations}
%Here, the only solvable parameter we consider for the $\tilde \bfp$ discretization 
%is the trajectory's final completion time $t_\text{f}$. 
%%%Here, 
%%%the trajectory's final completion time $t_\text{f}$ is  
%%%the only parameter we include in $\tilde \bfp$. % discretization 
%is 

%
%The linearization depends on the difference from the reference solution times the gradients 
%of equation~\eqref{eq:grad_ABE}. 
%, $\bar r(t)$ corresponds to an overall reference offset, 
%collecting the reference components into a single vector. 

%\begin{align}
%\frac{d \bfx(t(\tau))}{d\tau}
%=
%\frac{d \bfx(t(\tau))}{d t(\tau)}
%\frac{d t(\tau)}{d \tau}
%= t_f \frac{d \bfx(t)}{d t}
%\end{align}

%\begin{align}
%\frac{d \bfx(t)}{d\tau}
%=
%\frac{d \bfx(t)}{d t}
%\frac{d t}{d \tau}
%= t_f \frac{d \bfx(t)}{d t}
%\end{align}
%

%By setting $\bfp=[t_f]$ and $t = t_f \tau$, this becomes
%\begin{subequations}
%\begin{align}
%A_k &= \nabla_\bfx f(\bar \bfx(t_f k/N),\bar \bfu(t_f k/N))
%\\
%B_k &= \nabla_\bfu f(\bar \bfx(t_f k/N), \bar \bfu(t_f k/N))
%\\
%E_k &=  f(\bar \bfx(t_f k/N), \bar \bfu(t_f k/N)). 
%\end{align}
%\end{subequations}

%\begin{align}
%\bfx_{k+1}=
%\begin{bmatrix}
%A_k & B_k^- &B_k^+ & E_k 
%\end{bmatrix}
%\begin{bmatrix}
%\bfx_k 
%\\
%\bfu_k 
%\\
%\bfu_{k+1}
%\\
%\bfp 
%\end{bmatrix}
%+\bfr_k
%\end{align}

Using the stacked vectors from \eqref{eq:xup_stacked}, 
a discretized kinematic state equation, 
analagous to \eqref{eq:ABE_xup_r}, 
can be written as
\begin{align}
%\tilde\bfv
%0
%=
\underbrace{
\begin{bmatrix}
\tilde A & \tilde B & \tilde E
\end{bmatrix}
}_{K}
\begin{bmatrix}
\tilde \bfx
\\
\tilde \bfu
\\
\tilde \bfp 
\end{bmatrix}
+\tilde \bfr
=
0, 
\label{eq:K_xup_disc}
\end{align}
where 
$\tilde \bfr=\text{vec}(\bar\bfr_1, \ldots, \bar\bfr_N)$ is the stacked reference vector, 
and 
$K=[\tilde A, \, \tilde B,\,  \tilde E]$ is the overall kinematic matrix,   
with
\begin{subequations}
\begin{align}
\tilde A &= \begin{bmatrix}
A_1 & -1 & 0 & \cdots  & 0
\\
0 & \ddots & \ddots & & \vdots 
\\
\vdots & & \ddots & \ddots & 0
\\
0& \cdots &0& A_N & -1 
\end{bmatrix}
\\
\tilde B &= \begin{bmatrix}
B_1^- & B_1^+ & 0 & \cdots  & 0
\\
0 & \ddots & \ddots & & \vdots 
\\
\vdots & & \ddots & \ddots & 0
\\
0& \cdots &0& B_N^- & B_N^+
\end{bmatrix}
\\
\tilde E &= \begin{bmatrix}
E_1 
\\
\vdots 
\\
E_N
\end{bmatrix}. 
\end{align}
\label{eq:tilde_ABE}
\end{subequations}
%and $\tilde \bfr=(\bar\bfr_1, \ldots, \bar\bfr_N)$ represents the stacked reference vectors.

%\begin{align}
%\bfv_{1:N}
%=
%\begin{bmatrix}
%A_{N\times N} & B_{N\times N} & E_{N\times 1}
%\end{bmatrix}
%\begin{bmatrix}
%\bfx_{1:N} 
%\\
%\bfu_{1:N}  
%\\
%\bfp 
%\end{bmatrix}
%+\bfr_{1:N}
%\end{align}

Equation~\eqref{eq:tilde_ABE} is obtained by considering 
the discrete form of \eqref{eq:ABE_xup_r}, 
%which 
% represents
%%representing
%%which 
%%represents
%%corresponding to 
%%the transition
%%transitioning 
%a
%state transition 
% from index $k$ to  $k+1$, 
% and 
which 
 can be written
%and can be written, 
\begin{align}
\bfx_{k+1}
=
\begin{bmatrix}
A_k & B_k^- & B_k^+ & E_k 
%A_k & B_k^- & E_k 
\end{bmatrix}
\begin{bmatrix}
\bfx_k 
\\
\bfu_k 
\\
\bfu_{k+1}
\\
\bfp 
\end{bmatrix}
%+
%B_k^+ 
%\bfu_{k+1}
+\bfr_k. 
\label{eq:bfx_kp1_eq}
\end{align}
Expanding for each $k$,
and  rewriting 
\eqref{eq:bfx_kp1_eq}
in matrix form results in  
\eqref{eq:tilde_ABE}. 
Here, $A_k$ describes the control-free transition 
from state $\bfx_k$ to $\bfx_{k+1}$. 
A first-order hold is used to interpolate between control inputs $\bfu_k$ and $\bfu_{k+1}$, 
%which requires 
involving the
two control matrices $B_k^-$ and $B_k^+$, similar to the approach of Malyuta~et~al.~\cite{malyutadiscretization}. 
The matrix $E_k$ determines the effect of $\bfp$ on the state transition, 
and $\bfr_k$ describes the reference component. 
%corresponds to $\bar \bfr(t_\text{f}\,\tau_k)$, where $\tau_k=(k-1)/N$. 

%At each index $k$, 

%\begin{align}
%0 %\bfx_{k+1}
%=
%\begin{bmatrix}
%A_k & B_k^- &B_k^+ & E_k 
%%A_k & B_k^- & E_k 
%&-1 & B_k^+
%\end{bmatrix}
%\begin{bmatrix}
%\bfx_k 
%\\
%\bfu_k 
%\\
%\bfp 
%\\
%\bfx_{k+1}
%\\
%\bfu_{k+1}
%\end{bmatrix}
%+\bfr_k
%\end{align}

%\begin{align}
%\begin{bmatrix}
%1 & -B_k^+ 
%\end{bmatrix}
%\begin{bmatrix}
%\bfx_{k+1}\\
%\bfu_{k+1}
%\end{bmatrix}
%=
%\begin{bmatrix}
%%A_k & B_k^- &B_k^+ & E_k 
%A_k & B_k^- & E_k 
%\end{bmatrix}
%\begin{bmatrix}
%\bfx_k 
%\\
%\bfu_k 
%\\
%%\bfu_{k+1}
%%\\
%\bfp 
%\end{bmatrix}
%+\bfr_k
%\end{align}

%The matrices 
In the above equations, 
 $A_k$, $B_k$, $E_k$ and %vector 
 $\bfr_k$ 
%depend on discrete
involve
transition over the interval $k$ to $k+1$. 
%Evaluating these requires 
To evaluate these, 
we use
the state transition matrix 
$\Phi(t, t_0)$, 
%$\Phi_{t,t'}$, 
which can found by recursive numerical computation %by numerical integration from  
of the integral 
\begin{subequations}
\begin{align}
\Phi(t, t_0) &=
%I + \int_{\tau_k}^{\tau_{k+1}} A(t(\tau)) \Phi(\tau, \tau_k) \frac{dt}{d \tau} d\tau
I + \int_{t_0}^{t}  A(t) \Phi(t, t_0)  \,dt,
%\\&=
\label{eq:Phi_t_t0}
\end{align}
%using numerical methods, 
where 
$t$ is increased in small steps, starting from $t_0$, 
until $t=t_0+\Delta t$,
with an initial matrix
$\Phi(t_0, t_0)=I$. 

%To capture the impact of parameter $t_\text{f}$ on the state transition,  
For convenience, 
we replace \eqref{eq:Phi_t_t0} 
with $\Phi(\tau, \tau_{k})$, 
%where the domain is changed 
changing the domain
from $t$ to $\tau$, %={t/t_\text{f}}$, 
where $\tau_{k}=(k-1)/N$. 
This captures the impact of parameter $t_\text{f}$ on the state transition, 
scaling $t$, while keeping $N$ fixed \cite{reynolds2020real}. 
By direct substitution, we have  
%We define  $\Phi(\tau, \tau_{k})$ to represent 
%evaluating $\Phi(t,t_0)$, where $t=t_\text{f}\,\tau$ and $t_0=(k-1)/N$.  
\begin{align}
\Phi(\tau, \tau_{k}) &=
%I + \int_{\tau_k}^{\tau_{k+1}} A(t(\tau)) \Phi(\tau, \tau_k) \frac{dt}{d \tau} d\tau
%I + \int_{t_0}^{t_0+\Delta t_k}  A(t) \Phi(t, t_0)  \,dt
%\\&=
I + t_\text{f} \int_{\tau_k}^{\tau_{k+1}}  A(t_\text{f} \,\tau) \Phi(\tau, \tau_k)  \,d\tau,
\label{eq:Phi_tau_tau_k}
\end{align}
\end{subequations}
%$A(t(\tau))dt=A(t(\tau))dt$
where $t = t_\text{f}\, \tau$ and 
%$\tau = \frac{1}{t_f} t$,
 $dt = \frac{dt}{d \tau} d\tau = t_\text{f}\, d\tau$. 
%The discretization matrices depend on 
Now, 
using 
\eqref{eq:Phi_tau_tau_k}
to evaluate the
discretization matrices
  % equation~\eqref{eq:bfx_kp1_eq} 
(see Refs.~\cite{reynolds2020real,mao2018successive,malyutadiscretization}), we compute
\begin{subequations}
\begin{align}
A_k &= \Phi(\tau_{k+1}, \tau_k)
%end{align}
%\begin{align}
\\
B_k^- &= t_\text{f} A_k \int_{\tau_k}^{\tau_{k+1}} 
%\left(\frac{\tau_{k+1}-\tau}{\tau_{k+1}-\tau_{k}}\right)  
%\left(1-\frac{\tau-\tau_{k}}{\Delta \tau}  \right)
%\left(\frac{\tau_{k_1}-\tau}{\Delta \tau}  \right)
\Phi^{-1}(\tau, \tau_k)\,  B(t_\text{f}\, \tau) 
\,\eta_-(\tau)
\, 
d\tau
\\
B_k^+ &= t_\text{f} A_k \int_{\tau_k}^{\tau_{k+1}} 
%\left(\frac{\tau-\tau_{k}}{\tau_{k+1}-\tau_{k}}\right)  
%\left(\frac{\tau-\tau_{k}}{\Delta \tau}  \right)
\Phi^{-1}(\tau, \tau_k)\,  B(t_\text{f}\, \tau) 
\,\eta_+(\tau)
%(1-\eta(\tau))
\,
d\tau
\\
E_k &= t_\text{f} A_k \int_{\tau_k}^{\tau_{k+1}} 
\Phi^{-1}(\tau, \tau_k)\,  E(t_\text{f}\, \tau) 
\,
d\tau
\\
\bfr_k &= t_\text{f} A_k \int_{\tau_k}^{\tau_{k+1}} 
\Phi^{-1}(\tau, \tau_k)\,  \bar r(t_\text{f}\, \tau) 
\,
d\tau.
\end{align}
\label{eq:Ak_Bk_Ek_rk_int}
\end{subequations}
Here, 
$\eta_+(\tau) = \frac{\tau-\tau_{k}}{\Delta \tau}$ and 
$\eta_-(\tau) = \frac{\tau_{k+1}-\tau}{\Delta \tau}$
are used for linear interpolation of 
$B_k^-$ and $B_k^+$, where
$\Delta \tau = \tau_{k+1}-\tau_{k}$. 
To solve \eqref{eq:Ak_Bk_Ek_rk_int}, we evaluate \eqref{eq:grad_ABE}, yielding
\begin{subequations}
\begin{align}
A(t) %&= %\nabla_x f 
&=
\begin{bmatrix}
0&0&\cos\theta&0&-v \sin\theta&0\\
0&0&\sin\theta&0&-v \cos\theta&0\\
0&0&0&1&0&0\\
0&0&0&0&0&0\\
0&0&\frac{\tan\phi}{L}&0&0&0\\
0&0&0&0&\frac{v}{L}\sec^2\phi&0\\
0&0&0&0&0&0\\
\end{bmatrix}
\label{eq:A_t}
\end{align}
%
%A = zeros(input.nx,input.nx);
%A(1,3) = cos(theta);  %dx/dv
%A(1,5) = -v.*sin(theta);  %dx/dtheta 
%A(2,3) = sin(theta);  %dy/dv
%A(2,5) = v.*cos(theta);  %dy/dtheta 
%A(3,4) = 1.0; %dv/da ??
%A(5,3) = tan(phi)/L; %dtheta/dv ??
%A(5,6) = v.*sec(phi)^2/L; %dtheta/dphi ??
%%A()
%
%% if abs(v)<=0.1
%%     A(4,4)=-1; 
%% end
%
\begin{align}
B(t) %&= \nabla_u f 
&=
\begin{bmatrix}
0&0\\
0&0\\
0&0\\
1&0\\
0&0\\
0&1\\
\end{bmatrix}
\label{eq:B_t}
\end{align}
\begin{align}
E(t) &= %\frac{1}{t_\text{f}}
 f(\bfx(t), \bfu(t), \bfp).
 \label{eq:E_t}
\end{align}
%\begin{align}
%C(t) = 
%\end{align}
\label{eq:A_B_E_t}
\end{subequations}
%Here, 
The matrix
\eqref{eq:A_t} depends on %the
 states $v(t)$, $\theta(t)$ and $\phi(t)$ of  $\bfx(t)$ according to \eqref{eq:xt_ut}. 
%Thus, 
%At $t=t_\text{f}\,\tau$, 
%the states are computed as  
%the state vector is obtained from 
%The states 
%To evaluate %$A(t_\text{f}\,\tau)$, 
%In 
%equations
%To evaluate this, 
%For evaluation, 
To compute
\eqref{eq:Ak_Bk_Ek_rk_int} and 
\eqref{eq:A_B_E_t}, 
%we follow the approach of Malyuta et al.~\cite{malyutadiscretization}
%and Reynolds et al.~\cite{reynolds2020real}
a Runge-Kutta method (see Refs.~\cite{malyutadiscretization,reynolds2020real}) is used to perform
numerical integration of \eqref{eq:Phi_tau_tau_k} %is performed 
%using ,
in small steps. 
%For equation~\eqref{eq:A_B_E_t}, 
On the interval $\tau_k$ to $\tau_{k+1}$, 
%These are obtained, 
%For \eqref{eq:A_t}, 
%For equation~\eqref{eq:A_B_E_t}, 
the state vector 
%at $t=t_\text{f}\,\tau$ 
%is computed as 
is determined by 
%from 
%where
$\bfx(t_\text{f}\, \tau)=\Phi(\tau, \tau_k)\, \bfx_k$ and
% is used to evaluate $A(t_\text{f}\,\tau)$ in \eqref{eq:A_t}. 
%In equation~\eqref{eq:B_t}, the matrix $B(t)$ is constant, which simplifies \eqref{eq:Ak_Bk_Ek_rk_int}.  
%To evaluate \eqref{eq:E_t}, 
the
control vector
%$\bfu(t)$ 
is interpolated as 
$\bfu(t_\text{f}\,\tau)=\eta_+(\tau) \bfu_k + \eta_-(\tau) \bfu_{k-1}$.

%\subsection{Equations}
\section{Trajectory Optimization}
\label{sec:traj_opt}
%\begin{align}
%R(\theta)
%=\begin{bmatrix}
%cos(\theta) & -sin(\theta)\\
%sin(\theta) & cos(\theta)
%\end{bmatrix}
%\end{align}

%Sequential convex optimization involves solving the following sub-problem
%%%\begin{subequations}
%%\begin{align}
%%%\begin{equation}
%%%\begin{split}
%%{\bfz}^{(i+1)}
%%=
%%{\bfz}^{(i)}+
%%\underset{\Delta\bfz \in \scrK_i} %(\bfz^{(i)})}
%%{\text{argmin}} 
%%% &\quad 
%%\left[
%%% \varphi\left(\bfz^{(i)}+\Delta\bfz\right) 
%%\nabla_\bfz F ( \Delta\bfz)
%%%F\!\left(\bfz^{(i)}+
%%\nabla_\bfz F ( \Delta\bfz)
%%% F_\bfz\, \Delta\bfz
%%\right) 
%%+ \norm{\Delta\bfz}_\alpha
%% \right]
%%% \\&=
%%% \underset{\bfz \in \scrK_i+\bfz^(i)} 
%%%{\text{argmin}} 
%%%\left[
%%%\nabla_\bfz F ( \bfz^(i+1) - \bfz^(i))
%%%%F\!\left(\bfz^{(i)}+
%%%\nabla_\bfz F ( \Delta\bfz)
%%%% F_\bfz\, \Delta\bfz
%%%\right) 
%%%+ \norm{\Delta\bfz}_\alpha}
%%% \right]
%%\end{align}
%\end{subequations}

To obtain an optimal trajectory, we consider the linearized form of  equation 
\eqref{eq:K_xup_disc}
%consider the 
with the unknown vector variable $\bfz = \text{vec}(\tilde \bfx,\, \tilde \bfu,\, \tilde \bfp) $. 

In sequential convex optimization approaches \cite{yuan2015recent, malyuta2022convex}, 
%we approximate
a non-convex optimization problem, %with the form
represented as %in the form
\begin{subequations}
\begin{align}
\begin{split}
\underset{\bfz}{%
\text{minimize}}
&\quad
g(\bfz)
\\
\text{subject to}
&\quad
h(\bfz)\leq 0  
\end{split}
\label{eq:seq_conv_a}
\end{align}
%is approximated by the sequence
%can be approximated by 
can be approximated
with
a converging sequence of
convex sub-problems, given by
\begin{align}
\begin{split}
\bfz^{(i+1)}&=
\bfz^{(i)} +
\underset{\Delta\bfz \in \scrK_i}{%
\text{argmin}}
~
\nabla g\big(\bfz^{(i)}\big)^T \Delta \bfz
%+ \norm{\Delta \bfz}_\alpha^2
\\
&
\quad\quad
\text{subject to}
%\quad\!\!
%h(\bfz^{(i)})+\nabla h\left\!(\bfz^{(i)}\right)^T \Delta \bfz \leq 0 
~\nabla h\big(\bfz^{(i)}\big)^T \Delta \bfz \leq -h\big(\bfz^{(i)}\big)
,
%\nonumber
\end{split}
%\end{matrix}
\label{eq:seq_conv_b}
\end{align}%
\label{eq:seq_conv}%
\end{subequations}
where  $\bfz^{(i+1)}$ is the
%resulting
 solution %of 
of
%that results for 
%following 
%the 
%$i$-th 
iteration $i+1$.  
%In each iteration, 
Here, 
%In \eqref{eq:seq_conv_a},
the non-linear objective  $g(\bfz)$ and constraint  $h(\bfz)$  
of \eqref{eq:seq_conv_a}
are linearized 
in \eqref{eq:seq_conv_b}
by computing gradients $\nabla g(\bfz)$ and $\nabla h(\bfz)$ 
with respect 
to previous solution $\bfz^{(i)}$ (starting with initial solution $\bfz^{(0)}$). 

In \eqref{eq:seq_conv_b},
the unknown variable $\Delta \bfz$, which is solved,
%, which is solved in equation~\eqref{eq:seq_conv_b}, 
corresponds to the change in solution
$\bfz^{(i+1)}-\bfz^{(i)}$,
%thus
updating 
% and is used to update 
the result in each iteration. 
%Equation~\eqref{eq:seq_conv_b}
This %series
%solution sequence
sequence of solutions
% converges 
will converge
 to a local minimum when 
$\Delta \bfz$ is confined to a suitable trust region $\scrK_i$ 
%surrounding each previous
anchored at 
$\bfz^{(i)}$\cite{yuan2015recent}. 
%the previous solution. 
%For simplicity,
%to implement a penalized trust region, 
A penalized trust region for $\scrK_i$ can be implemented  
% we %limit  
%achieved here with the
using an $\alpha$-weighted norm
$\norm{\Delta \bfz}_\alpha$, 
where $\alpha$ specifies a weighting between elements of $\Delta \bfz$. %, %. This penalizes solutions
%In this way, $\alpha$ permits scaling of   to scaled the relative we
%promoting solutions near the
%%that deviate from the 
%% deviating from the 
% previous iteration's result.  
The weighting %allows normalizing 
%normalizes
%can be
is
selected to normalize
each element %of $\Delta \bfz$ 
to its range of valid parameters, 
%which scales
%the
scaling the
step-sizes along individual dimensions for improved convergence. 
 
%The form of  

%Equality constraints %can be %inherent to
%are
%a natural extension of 
%equation~\eqref{eq:seq_conv},  %can support
%corresponding to
%two identical 
%inequality constraints of opposite sign. %, % are used, % yield an equality,  

%inequality constraints %, 
%by adding a positive valued slack variable to exceed the equality in one direction, while 
%%modifying the objective function 
%minimizing the slack variable's norm in the objective. 
%For clarity, 

 %in the objective. %For convenience  

%%%
%%%\begin{subequations}
%%%\begin{align}
%%%\bfz^{(i+1)}&=
%%%\bfz^{(i)} +
%%%\underset{\Delta\bfz}{%
%%%\text{argmin}}
%%%%&~
%%%~\nabla g\big(\bfz^{(i)}\big)^T \Delta \bfz
%%%+ \norm{\Delta \bfz}_\alpha^2
%%%%\\
%%%%\quad\quad\quad\,\,
%%%%\text{subject to}
%%%%%\quad\!\!
%%%%%h(\bfz^{(i)})+\nabla h\left\!(\bfz^{(i)}\right)^T \Delta \bfz \leq 0 
%%%%&~\nabla h\big(\bfz^{(i)}\big)^T \Delta \bfz \leq -h\big(\bfz^{(i)}\big) 
%%%%\\&=
%%%\end{align}
%%%which is equal to 
%%%\begin{align}
%%%\bfz^{(i)} +
%%%\bfr^{(i)} +
%%%\underset{\bfz}{%
%%%\text{argmin}}
%%%%&~
%%%~\nabla g\big(\bfz^{(i)}\big)^T 
%%%\,\bfz
%%%+ \norm{\bfz- \bfz^{(i)}}_\alpha^2 
%%%\end{align}
%%%\end{subequations}
%%%where $\bfr^{(i)}=-\nabla g\big(\bfz^{(i)}\big)^T \, \bfz^{(i)}$.

%%To solve a trajectory we compute
%The optimal trajectory is %obtained by repeated solving of
%solved in sequence,
To solve an optimal trajectory, we 
use equation~\eqref{eq:seq_conv} 
%solved using equation~\eqref{eq:seq_conv}, 
 with the $i$-th iteration 
% given 
specified
 by 
%iterations in the sequence
\begin{subequations}
\begin{samepage}
\begin{align}
%\begin{equation}
%\begin{split}
\begin{bmatrix}
\tilde\bfx^{(i+1)}
\\
\tilde\bfu^{(i+1)}
\\
t_\text{f}^{(i+1)}
\end{bmatrix}
=
%\underset{(\tilde\bfx, \tilde\bfu, t_f)}
%\underset{\tilde\bfx, \tilde\bfu, t_\text{f}, \rho}
\underset{\tilde\bfx,\, \tilde\bfu,\, t_\text{f}}
%{\text{minimize}} 
{\text{argmin}} 
 ~ t_\text{f} 
% +\rho 
 + \varphi(\tilde \bfx, \tilde \bfu)  %, \tilde \bfv) 
+
\norm{
\begin{matrix}
\tilde\bfx-\tilde\bfx^{(i)}
\\
\tilde\bfu-\tilde\bfu^{(i)}
\\
t_\text{f}- t_\text{f}^{\,(i)}
\end{matrix}
}_\alpha^2
% + \varphi(\tilde \bfx, \tilde \bfu)  
%\\
\label{eq:min_traj_a}
\end{align}
\begin{align}
%\\
\text{subject to} &\quad  
%\tilde
 K 
\begin{bmatrix}
\tilde \bfx
\\
\tilde \bfu
\\
t_\text{f}
%\tilde \bfp 
\end{bmatrix}
+
\tilde \bfr %^{(i)}
=
0 
\label{eq:min_traj_b}
%\tilde \bfv 
\\
&\quad  M \tilde \bfx + \bfb \leq 0
\label{eq:min_traj_c}
\\
\begin{split}
\label{eq:min_traj_d}
&\quad  
%%%%(\bfx_1, \bfu_1) = (\bfx_\text{init}, \bfu_\text{init})
\bfx_1 = \bfx_\text{init},\quad \bfu_N=\bfu_\text{final}
\\
&\quad  
%(\bfx_N, \bfu_N) = (\bfx_\text{final}, \bfu_\text{final})
\bfu_1=\bfu_\text{init}, \quad \bfx_N = \bfx_\text{final}
%%%\begin{bmatrix}
%%%\bfx_1 \\ \bfu_1
%%%\end{bmatrix}
%%%=
%%%\begin{bmatrix}
%%%\bfx_\text{init} \\ \bfu_\text{init}
%%%\end{bmatrix}
%%%,\quad
%%%\begin{bmatrix}
%%%\bfx_N \\ \bfu_N
%%%\end{bmatrix}
%%%=
%%%\begin{bmatrix}
%%%\bfx_\text{final} \\ \bfu_\text{final}
%%%\end{bmatrix}
\end{split}
\\
%&\quad  (\bfx_\text{min}, \bfu_\text{min}) 
%		\leq (\tilde \bfx, \tilde \bfu) 
%		\leq (\bfx_\text{max}, \bfu_\text{max})
\begin{split}
\label{eq:min_traj_e}
&\quad  
%%%\begin{bmatrix}
%%%\bfx_\text{min}\\ \bfu_\text{min}\\t_{f,\text{min}}
%%%\end{bmatrix} 
%%%		\leq
%%%\begin{bmatrix}
%%%\tilde \bfx
%%%\\ 
%%%\tilde \bfu
%%%\\
%%%t_f 
%%%\end{bmatrix} 
%%%\leq
%%%\begin{bmatrix}
%%%\bfx_\text{max}\\ \bfu_\text{max}\\t_{f,\text{max}}
%%%\end{bmatrix} 
%
\begin{bmatrix}
\bfx_\text{min}\\ \bfu_\text{min}
\end{bmatrix} 
		\leq
\begin{bmatrix}
\bfx_k
\\ 
 \bfu_k
\end{bmatrix} 
\leq
\begin{bmatrix}
\bfx_\text{max}\\ \bfu_\text{max}
\end{bmatrix} 
%,  k=1\ldots N
%1\leq k \leq N
%
%\bfx_\text{min}\leq \bfx_k \leq \bfx_\text{max}, \quad k=1\ldots N
%\\&\quad
%\bfu_\text{min}\leq \bfu_k \leq \bfu_\text{max}, \quad k=1\ldots N
\end{split}
\\&\quad
t_\text{f}^\text{min}\leq t_\text{f} \leq t_\text{f}^\text{max}
\label{eq:min_traj_f}
%
%		\leq (\tilde \bfx, \tilde \bfu) 
%		\leq (\bfx_\text{max}, \bfu_\text{max})
%\\
%&\quad  
%\norm{%\bfalpha\cdot
%%(\tilde\bfx-\bar\bfx^{(i)},\tilde\bfu-\bar\bfu^{(i)},  t_f-\bar t_f^{(i)})
%\begin{matrix}
%\alpha_x\left(\tilde\bfx-\bar\bfx^{(i)}\right)
%\\
%\alpha_u\left(\tilde\bfu-\bar\bfu^{(i)}\right)
%\\
%\alpha_t\left(t_f-\bar t_f^{(i)}\right)
%\end{matrix}
%%%\norm{
%%%\begin{matrix}
%%%\tilde\bfx-\bar\bfx^{(i)}
%%%\\
%%%\tilde\bfu-\bar\bfu^{(i)}
%%%\\
%%%t_\text{f}-\bar t_\text{f}^{\,(i)}
%%%\end{matrix}
%%%}_\alpha
%%%\leq \rho		
%\end{split}
%\end{equation}
\end{align}
\label{eq:min_traj}
\end{samepage}
\end{subequations}
The first term of \eqref{eq:min_traj_a} minimizes the time to reach final state $t_\text{f}$. 
The second term $\varphi(\tilde \bfx, \tilde\bfu)$ is used to regularize the state and control variables. 
This promotes smoothness and minimizes energy used in the trajectory, and also results in faster convergence.  
The third term $\norm{\cdot}_\alpha$ penalizes results 
far from the previous solution, with each variable weighted by vector $\alpha$. 
Equation~\eqref{eq:min_traj_b} is an equality constraint 
to ensure valid dynamics
corresponding to equation~\eqref{eq:K_xup_disc}. 
Collision avoidance is implemented with 
\eqref{eq:min_traj_c},
%where $M \tilde \bfx + \bfb \leq 0$ ensures
%%ensuring 
%trajectory waypoints are confined to halfspaces defined by nearby objects, 
as described in Section~\ref{sec:collision_avoid}. 
Equation~\eqref{eq:min_traj_d} is used to fix the position of the 
initial and final waypoints and control variables. 
Equation~\eqref{eq:min_traj_e} ensures that each state and control variable is bounded between minimum and maximum values
 for $k=1\ldots N$. 
Eqaution~\eqref{eq:min_traj_f} sets limits on $t_\text{f}$. % the final time. 
 
%, 
%where $M \tilde \bfx + \bfb$ ensures trajectory waypoints are 
%do not .   
  
%\eqref{eq:min_traj_a}, 
%\eqref{eq:min_traj_b}, 
%%\eqref{eq:min_traj_c}, 
%\eqref{eq:min_traj_d}, 
%\eqref{eq:min_traj_e}, 
%\eqref{eq:min_traj_f}, 

%
%\begin{align}
%%r(t) &= f(\bar \bfx(t), \bar \bfu(t), \bar t) - A(t)\bar \bfx(t) - B(t)\bar \bfu(t) - E( t) \bar t
%r(t) &= 
%f(\bar \bfx(t), \bar \bfu(t), \bar t) - 
%\begin{bmatrix}
%A(t)\\B(t)\\ E(t)
%\end{bmatrix}
%\begin{bmatrix}
%\bar \bfx(t) &  
%\bar \bfu(t) &
%\bar t
%\end{bmatrix}
%%\begin{bmatrix}
%%A(t)\bar \bfx(t) - B(t)\bar \bfu(t) - E(\bar t) t
%%\end{bmatrix}
%\end{align}
%
%\begin{align}
%\frac{d x(t)}{dt}&= f(x(t), u(t), t)	
%\\&approx 
%\end{align}

\section{Collision Avoidance}

\label{sec:collision_avoid}

In this section, we formulate the collision avoidance constraints of equation~\eqref{eq:min_traj_c}. 
%First, we formulate a %collision avoidance 
This involves specifying one
constraint for each waypoint with respect to each obstacle. 
%it indicates penetration into the obstacle, whereas a negative \( f \) indicates a safe distance. 
%Additionally, %we compute 
In each iteration of equation~\eqref{eq:min_traj}, it is efficient to compute 
 $d_{kj}$, 
%corresponds to 
%which represents 
the closest distance from 
the $j$-th obstacle to 
the polygonal boundary surrounding waypoint $x_k$. %, remains positive. 

If $d_{kj}$ is positive, this 
indicates a safe distance to the vehicle,  
whereas a negative value indicates penetration into the obstacle. 
An individual constraint is given by 
\begin{align}
d_{kj} + \nabla d_{kj}^T \bfx_k   \geq 0,   
\label{eq:dkj}
\end{align}
where the gradient 
$\nabla d_{kj}$, 
%the gradient of $d_{kj}$ 
%computed
%which is
taken
with respect $\bfx_k$, %which 
represents how 
variations in the state variables %\( x \), \( y \), and \(\theta\) 
impact distance to the object, as described below.

For collision avoidance, 
using a polygonal  boundary %for the vehicle 
%is used to fill 
%is used
%because 
%to fill
%to 
permits filling of
gaps that result from sampling and curvature of successive waypoints, which is apparent in Figures~\ref{fig:fig2} and \ref{fig:fig5}, where spacing between adjacent rectangles creates a \enquote{sawtooth} pattern. 
% rather than
% a smooth curve of 
% the trajectory's envelope. 
 To fill in these gaps, polygon $k$ is %defined as
taken as 
the convex hull of two rectangles at  adjacent waypoints $\bfx_k$ and $\bfx_{k+1}$, %which are defined by the vehicle's 
%defined by 
%which are
%where each is 
% defined according 
%which are
transformed
according 
 to 
the vehicle's %c%urrent
position 
%$(x_k, y_k)$
and 
orientation. 
%$\theta_k$. 
%and 
%defined by 
%with dimensions specified by 
%its constant 
%using rectangular
As depicted in Figure~\ref{fig:fig1}, 
the vehicle has length $L+L_f+L_r$
and width $L_w$, where $L_f$ and $L_r$ are the front and rear overhang and $L$ is the wheel base.   
%Regularization to limit the distance between adjacent waypoints  

The vehicle polygon consists of a list of edges and vertices %surrounding $\bfx_k$. 
that surround waypoint $k$. 
%Each obstacle is a rectangle,
The $j$-th obstacle is represented by a rectangle,
%with center position, length, width and orientation parameterized by 
%$(x_0, y_0, a_j, b_j, \gamma_j)_j$. 
%$(x_j, y_j, a_j, b_j, \psi_j)$. 
with center position
$(x_j, y_j)$, 
length $a_j$, width $b_j$ and orientation $\psi_j$. 
% centered at $(x_0, y_0)^{j}$, with length $a_j$, width $b_0$ and orientation $\theta_0$. 
To compute the closest point in an efficient way, 
the polygon is rotated into the $j$-th
obstacle's 
principal coordinate frame, according to 
\begin{align}
\begin{bmatrix}
g_x\\g_y
\end{bmatrix}
=
%\left(
%\begin{bmatrix}
%{a_j}&0\\
%0&{b_0}\\
%\end{bmatrix}
%^{-1}
%\!\!
R^{-1}(\psi_j)
%\right)^{-1}
\begin{bmatrix}
p_x-x_j\\p_y-y_j
\end{bmatrix},
\end{align}
where $(p_x,p_y)$ is a polygon vertex. %, %and the %. The 
%and 
%$R(\theta)$ is 
%the 
%with 
The
rotation matrix %is %given by 
is
\begin{align}
R(\psi_j) = 
\begin{bmatrix} 
\cos(\psi_j) & -\sin(\psi_j) 
\\ 
\sin(\psi_j) & \cos(\psi_j) 
\end{bmatrix}. 
\end{align}

When the polygon does not intersect the rectangle,  
the closest point
$(\tilde g_x,\tilde g_y)$ %is set to 
%represents
%%the point on the polygon closest to the rectangle, 
%, 
is
found by identifying the minimum distance between the polygon's edges and the rectangle's sides using separating-axis projections. 
% closet distance from any edge 
%to the rectangle. 
However, if any vertices are found within the rectangle
(i.e., $|g_x| < a_j$ and $|g_y|< b_j$), then
$(\tilde g_x,\tilde g_y)$ 
%this
is set to their average position. 
Otherwise, if an edge has two intersections with the rectangle, 
%it is set to their midpoint. 
their midpoint is used. 
%the midpoint of these is used. 

%Using %the closest point, 
%this 
%The constraint of 
Equation~\eqref{eq:dkj} is specified, 
using the closest point, according to  
%
%If outside of the rectangular object (i.e., $|g_x|\geq a_j$ or $|g_y|\geq b_j$) 
%then the closet distance is given by 
%%$d_{kj}=\norm{(d_{kj}^x, d_{kj}^y)}$, 
%$d_{kj}=\sqrt{(d_{kj}^x)^2+ (d_{kj}^y)^2}$, 
%where 
%$d_{kj}^x=\max(0, g_x-a_0, -g_x-a_0)$
%and 
%$d_{kj}^y=\max(0, g_y-b_0, -g_y-b_0)$. 
\begin{subequations}
\begin{align}
d_{kj}=
\begin{cases}
\sqrt{(d_{kj}^x)^2+ (d_{kj}^y)^2}, 
%\quad
%\norm{
%(d_{kj}^x, \, d_{kj}^y)
%%\end{bmatrix}
%},
&|\tilde g_x|>a_0 \text{ or } |\tilde  g_y|>b_0, 
\\
\bfn^T %\cdot 
\begin{bmatrix}
%g_x - a_j\, \text{sign}(g_x)\\
%g_y - b_j\, \text{sign}(g_y)
\tilde g_x - a_j\, s_x\\
\tilde g_y - b_j\, s_y
\end{bmatrix},
&\text{otherwise }
\end{cases}
\end{align}
where 
%Here, 
$d_{kj}^x=\max(0, \tilde g_x-a_j, -\tilde g_x-a_j)$ and
$d_{kj}^y=\max(0, \tilde g_y-b_j, -\tilde g_y-b_j)$. 
Here, 
for brevity, we set
$s_x = \text{sign}(\tilde g_x)$ and 
$s_y = \text{sign}(\tilde g_y)$. 
The gradient is computed as
\begin{align}
\nabla d_{kj}
%&=
%\left(
%\frac{\partial}{\partial x},
%\frac{\partial}{\partial y},
%\frac{\partial}{\partial v},
%\frac{\partial}{\partial a},
%\frac{\partial}{\partial \theta},
%\frac{\partial}{\partial \phi}
%\right)
%d_{kj} \nonumber
%\\
&=
%(
[
\tilde n_1, \tilde n_2, 0, 0, 
\tilde n_2 \cos(\theta_k)-
\tilde n_1 \sin(\theta_k), 0]^T,
%\begin{bmatrix}
%\tilde n_1\\
% \tilde n_2\\ 
% 0 \\ 
%0 \\
%\tilde n_2 \cos(\theta_k)- \tilde n_1 \sin(\theta_k) \\
% 0
%\end{bmatrix}
\end{align}
\label{eq:dkj_nabla_dkj}
\end{subequations}
where
$\theta_k$
%$\theta_k = (\bfx_k)_5$
is the vehicle orientation at waypoint $\bfx_k$. 
The
direction vector $\tilde \bfn=(\tilde n_1,\tilde n_2)$ is
\begin{align}
\tilde \bfn = R(\psi_j) \, \bfn, 
\end{align}
where
\begin{align}
\bfn = 
\begin{cases}
\frac{(\tilde g_x, \tilde g_y)}{\norm{(\tilde g_x, \tilde g_y)}} , 
%\widehat{(g_x, g_y)} , 
&
|\tilde g_x|>a_j \text{ or } |\tilde g_y|>b_j, 
\\
%\frac{1}{\sqrt 2} (s_x, s_y), 
\frac{(s_x, s_y)}{\norm{(s_x, s_y)}} , 
%\widehat{(s_x, s_y)} , 
& \max(|( |\tilde g_x| - a_j)|,\, 
|( |\tilde g_y| - b_j)|)
   \leq \epsilon 
%   \text{ and } 
%|( |g_y| - b_j) | \leq \epsilon
\\
%(\text{sign}(g_x), 0), 
(s_x, 0), 
& 
%s_x = s_y \text{ and } 
\left|\frac{\tilde g_x}{a_j} \right|  
\leq
\left|\frac{\tilde g_y}{b_j} \right|  
\\
(0, s_y), 
& \text{otherwise.}
\end{cases}
\label{eq:v}
\end{align}

In equation~\eqref{eq:v},
when there is no collision, 
 $\bfn$ represents the direction of the closest point to the obstacle. 
%  when 
%they are not in contact. 
%If the polygon would  a collision, 
If a collation will occur, 
%the 
constraint
\eqref{eq:dkj}
 becomes
%will become
  active, %if the vehicle collides with an obstacle, 
pushing the waypont in direction $\bfn$ for the next iteration. 
%there is no overlap. Otherwise, 
%However, 
To achieve effective performance, 
if 
$(\tilde g_x, \tilde g_y)$ is inside the obstacle,
%the closest point is
%the constraint 
when it falls 
%if it is 
 within distance $\epsilon$ of a corner, 
%equation~\eqref{eq:v} orients 
%the direction
$\bfn$
%pushes away
is
%will be
 oriented
 at a $45^\circ$ angle; 
otherwise, 
%the constraint 
%it
$\bfn$ will be oriented to 
%pushes
%will
 push away the waypoint
 in a 
 direction
%is oriented
perpendicular to
%  towards 
  the rectangle's closest side. 
In this manner, by snapping to angles in increments of $45^\circ$ relative to $\psi_j$, 
the gradient $\nabla d_{kj}$ remains more stable between successive iterations, 
which 
reduces oscillation, % and 
leading to faster convergence.

The collision avoidance constraints of equation~\eqref{eq:min_traj_c} are obtained as
\begin{subequations}
\begin{align}
\begin{split}
M&=\text{diag}(M_1, \ldots, M_N), 
\\
%&
\bfb&=\text{vec}(\bfb_1, \ldots, \bfb_N),   
\end{split}
\end{align}
%where  matrices $M_k$ are on the block diagonal of $M$, 
%and column vectors $\bfb_k$ are concatinated into $\bfb$, 
%where the matrices $M_k$ and column vectors $\bfb_k$ are defined as 
where each sub-matrix $M_k$ and column vector $\bfb_k$, for $k=1\ldots N$, is defined as 
%. These
%which
% are defined as
\begin{align}
\begin{split}
%M_k &= (\nabla d_{k1}^T, \ldots, \nabla d_{kJ}^T),
M_k &= [\nabla d_{k1}, \ldots, \nabla d_{kJ}]^T,
\\
%&
%\bfb_k &= (d_{k1}, \ldots, d_{kJ}),
\bfb_k &= [d_{k1}, \ldots, d_{kJ}]^T. 
\label{eq:Mkbk} 
\end{split}
\end{align}
\end{subequations}
%where %the $j$-th row % of $M_k$ and $b_k$ 
%such that a 
Thus, 
%a row
row $j$ 
in \eqref{eq:Mkbk}
%(of $M_k$ and $b_k$)
%is the constraint 
corresponds to equation~\eqref{eq:dkj_nabla_dkj}
for the $j$-th 
obstacle %$j$ %constraint 
%of obstacle $j$
and $k$-th waypoint. % $k$. 

\section{Blind Spot Avoidance}

\label{sec:blind_spot}

In general, it is undesirable for an autonomous vehicle to move into an area that has been left unchecked by sensors 
without confirming that no obstacle is present. 
While sensor placement provides sufficient coverage 
to detect obstacles
in most scenarios, 
potential issues can arise due to 
sensor latency, sensor blockage and high steering angle.
In some situations, there is potential for sides of the host vehicle 
to enter blind-spots 
when it is turning. 
As well, 
when the host vehicle 
or 
obstacle
are moving, 
%is moving, or obstacles are non-stationary, 
sensors may have insufficient time to distinguish a sudden detection from background noise, despite proper coverage. 
Moreover, sensor near-field zones may have limitations in detecting obstacles or estimating their distances. 

In this section we propose constraints so the generated trajectory will avoid passing through blind spots. 
The approach involves using a known 
spatial map of the sensor 
field-of-view, which is used to maintain a map of regions that remain unchecked at each waypoint in the trajectory. 
In each iteration of equation~\ref{eq:min_traj}, 
modified bounds constraints are computed to prevent the vehicle from traversing through the unchecked region.

In our approach, the sensor field-of-view is modeled as an image mask, %$V(\bfq)$, according to 
which is given by
\begin{align}
	V(\bfq) =
	\begin{cases}
	0, & \text{if $\bfq$ is a blind spot,}
	\\
	1, & \text{if $\bfq$ is visible,}
	\end{cases}
\end{align}
where $\bfq=(q_x, q_y)$ is a coordinate in the host vehicle's local frame. 
%In the global coordinate frame $\bfr=(r_x,r_y)$,
The unchecked region for waypoint $k$  is
\begin{subequations}
\begin{align}
	U_k(\bfr) = 1-C_k(\bfr), 
\end{align}
where the checked region $C_k(\bfr)$ is formed by transforming 
$V(\bfq)$ relative to waypoint $k$, 
and computing a union with the previous checked region $C_{k-1}(\bfr)$. %, according to 
Thus, 
\begin{align}
	C_k(\bfr) = C_{k-1}(\bfr) \, \cup \,  V\!\left(R_{\theta_k} (\bfr  - \bfr_k)\right).  
\end{align}
\end{subequations}
Here, 
$\bfr=(r_x,r_y)$ is a coordiante in the global frame, 
$\bfr_k=(x_k,y_k)$ is the vehicle's position and  $R_{\theta_k}=R(\theta_k)$ is a rotation 
%matrix, and $\theta_k$ is the waypoint orientation. 
by vehicle orientation  $\theta_k$. 
The initial $C_0$ is set to 1 in a region surrounding the host vehicle, and 0 elsewhere. 
%\begin{align}
%	U_k(\bfr) = 1- (1-U_{k-1}(\bfr)) \, \cup \,  V\!\left(R_{\theta_k} (\bfr  - \bfr_k)\right), 
%\end{align}

To compute modified bounds constraints, specific increases or decreases to $x_k$, $y_k$ or $\theta_k$ 
%by 
%$\Delta x_k$, $\Delta y_k$ or $\Delta \theta_k$ 
%could result in collision with 
%will cause the vehicle to enter unchecked region
%could %cause the vehicle to 
%can
are 
% be
  found that
%make the vehicle 
would cause the host vehicle's boundary polygon to
intersect  
$U_k(\bfr)$. 
%the unchecked region. 
%By determining an upper or lower ranges,
%Thus,
%to prevent entering the unchecked region, 
In particular, 
% we can modify 
 constraint~\eqref{eq:min_traj_e}
 is modified so 
% , such that 
$\bfx^\text{min}_k$  and $\bfx^\text{max}_k$ 
depend on $k$. The new limits are given by %, such that
%are 
%according to 
%specific to waypoint $k$, according to 
%for $\Delta x_k$, $\Delta y_k$ or $\Delta \theta_k$
\begin{subequations}
\begin{align}
\begin{split}
x^\text{min}_k&=\max(x^\text{min}, x_k-\Delta x_k^-)
%&x^\text{max}_k&=\min(x^\text{min}, x_k+\Delta x_k^+)
%\nonumber
\\
y^\text{min}_k&=\max(y^\text{min}, y_k-\Delta y_k^-)
%&y^\text{max}_k&=\min(y^\text{min}, y_k+\Delta y_k^+)
%\nonumber
\\
\theta^\text{min}_k&=\max(\theta^\text{min}, \theta_k-\Delta \theta_k^-)
%&\theta^\text{max}_k&=\min(\theta^\text{min}, \theta_k+\Delta \theta_k^+)
\end{split}
\end{align}
and
\begin{align}
\begin{split}
x^\text{max}_k&=\min(x^\text{max}, x_k+\Delta x_k^+)
%\nonumber
\\
y^\text{max}_k&=\min(y^\text{max}, y_k+\Delta y_k^+)
%\nonumber
\\
\theta^\text{max}_k&=\min(\theta^\text{max}, \theta_k+\Delta \theta_k^+)
\end{split}
\end{align}
\end{subequations}
where 
%changes of 
$\Delta x_k^\pm$, $\Delta y_k^\pm$ and $\Delta \theta_k^\pm$
are the smallest displacements 
for the vehicle polygon
to reach a non-zero in $U_k(\bfr)$.  
% would collide with the unchecked region. 

%The unchecked region $U_k(\bfq)$ for waypoint $k$ is
%\begin{align}
%	U_k(\bfr) = U_{k-1}(\bfr)  T_k\left\{ V\left(R(\theta_k) \bfr  - \begin{bmatrix}
%	x_k \\ y_k
%	\end{bmatrix}\right)\right\}
%\end{align}

\section{Results and Discussion}

\label{sec:results}

%\begin{figure}[!htb]
%    \centering
%    \begin{minipage}{.5\textwidth}
%        \centering
%        \includegraphics[width=0.3\linewidth, height=0.15\textheight]{prob1_6_2}
%        \caption{$dt=0.1$}
%        \label{fig:prob1_6_2}
%    \end{minipage}%
%    \begin{minipage}{0.5\textwidth}
%        \centering
%        \includegraphics[width=0.3\linewidth, height=0.15\textheight]{prob1_6_1}
%        \caption{$dt =$}
%        \label{fig:prob1_6_1}
%    \end{minipage}
%\end{figure}

%\begin{figure*}[t]
\begin{figure}[b!]
  \centering
%  \begin{minipage}[b]{.48\textwidth}
      \centering
      \includegraphics[width=0.7\columnwidth]{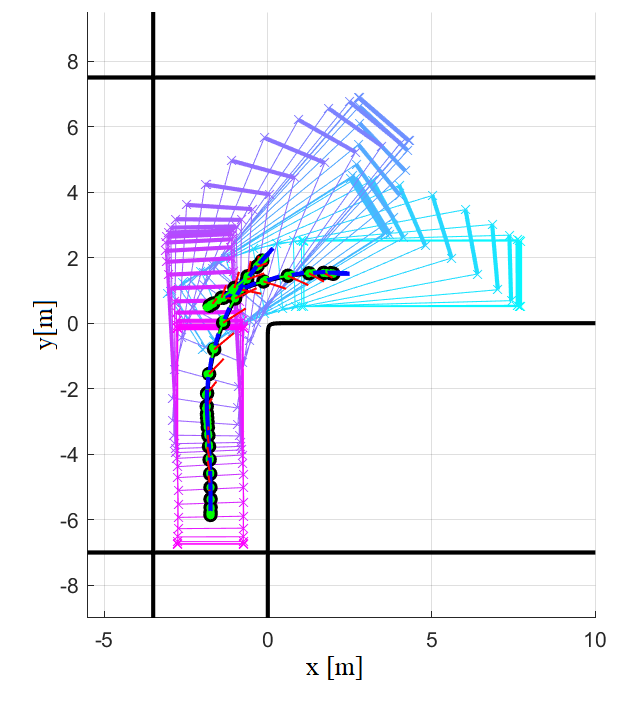}  
       
      \includegraphics[width=0.8\columnwidth]{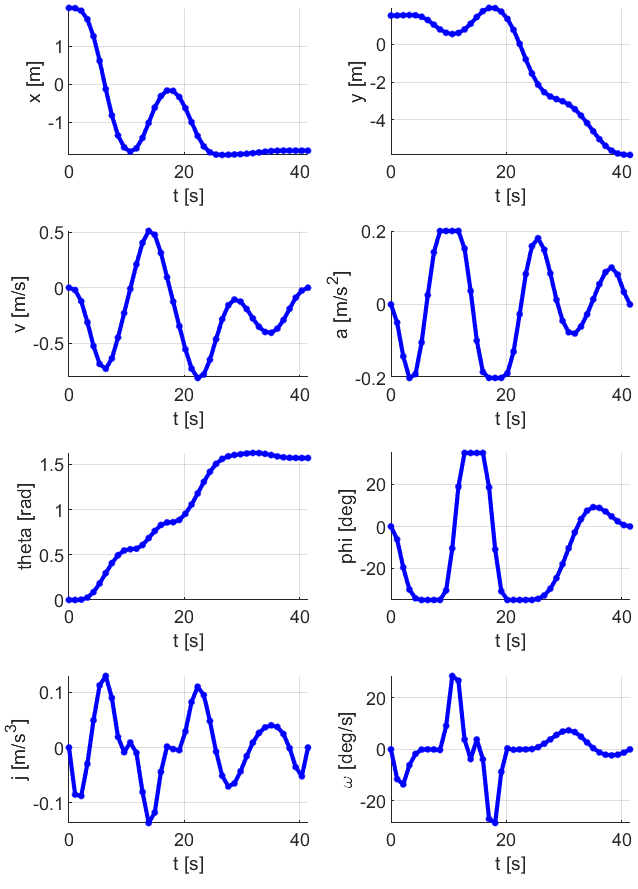}
%      \includegraphics[width=0.7\columnwidth]{figs/solution_out_01a.png}
	  %\caption{Rear-in perpendicular parking maneuver. }
	      \caption{Rear-in perpendicular parking maneuver. 
%	      The vehicle's path is shown with each waypoint represented as a rectangle, transitioning from cyan (initial position) to magenta (final position). 
	      The plots show state and control variables over time, demonstrating adjustments for
	       a 
	       smooth and 	       
	       collision-free trajectory into a perpendicular slot.
The vehicle boundary transitions from cyan (initial position) to magenta (final position).     
	      }	      
  	  \label{fig:fig2}
% \end{minipage}%
% \hspace{0.04\textwidth}% 
% \begin{minipage}[b]{0.48\textwidth}
\end{figure}
\begin{figure}[t!]
   		\centering
  		\includegraphics[width=0.8\columnwidth]{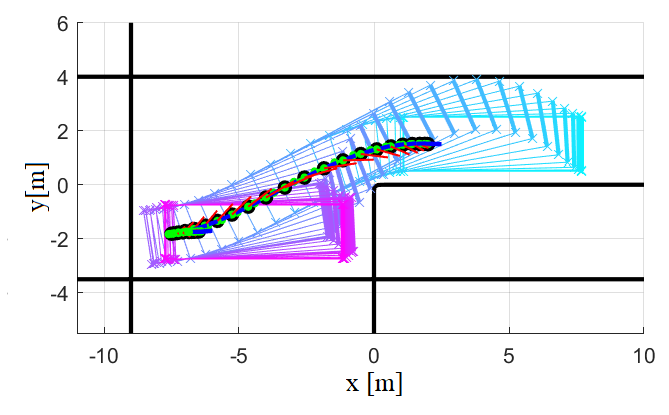}

  		\includegraphics[width=0.8\columnwidth]{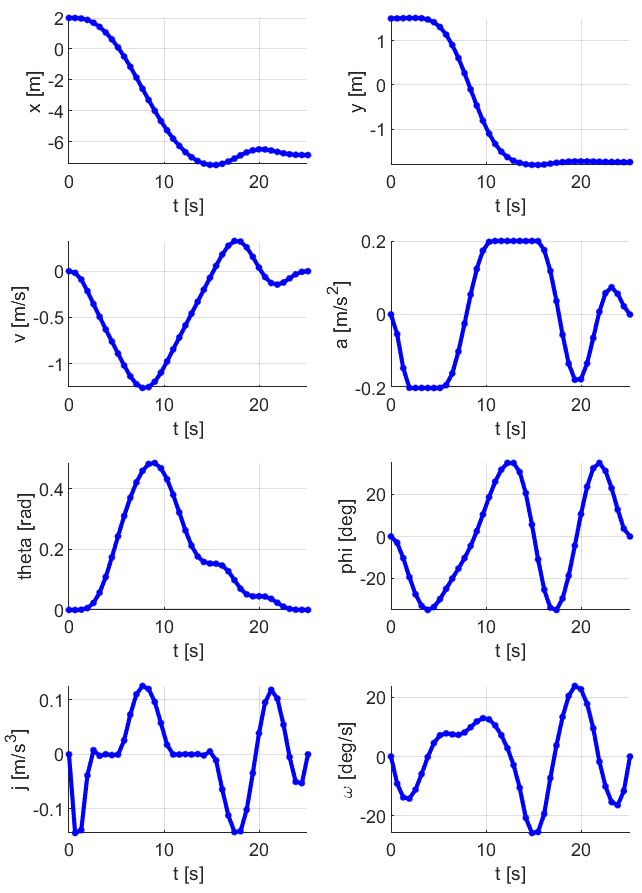}
	  	\caption{Rear-in parallel parking maneuver.  The plots illustrate adjustments in position, velocity, and steering to align parallel to the curb and fit into the parking slot. %, avoiding obstacles. 
%	  	The vehicle's outline transitions from cyan to magenta.
    The vehicle boundary transitions from cyan (initial position) to magenta (final position). 	  	
	  	}
%   	\caption{Rear-in parallel parking maneuver. 
%   	The vehicle's path is shown with each waypoint represented as a rectangle, transitioning from cyan (initial position) to magenta (final position). The plots below show the state and control variables over time.
%   	}
  		\label{fig:fig3}
% \end{minipage}
\end{figure}

%%%\begin{figure}[b]
%%%  \centering
%%%  \includegraphics[width=0.8\columnwidth]{figs/solution_out_02b.png}
%%%  
%%%  \includegraphics[width=0.8\columnwidth]{figs/solution_out_02a.png}
%%%  \caption{Rear-in parallel parking maneuver.}
%%%  \label{fig:fig2}
%%%\end{figure}

%\begin{figure}[t]
%  \centering
%  \includegraphics[width=0.5\textwidth]{figs/gui1.png}
%  \caption{caption}
%  \label{fig:fig2}
%\end{figure}

%%%\begin{figure*}[bt!]
%%%  \centering
%%%  \begin{minipage}[b]{.48\textwidth}
%%%      \centering
%%%%	  \includegraphics[width=1.0\columnwidth]{figs/gui_out1.png}
%%%	  \includegraphics[width=.70\columnwidth]{figs/gui_out1.png}
%%%
%%%	  \includegraphics[width=1.0\columnwidth]{figs/gui_out1a.png}
%%%    \caption{Parallel park out maneuver.  The plots show how the vehicle maneuvers out of the parallel parking slot by adjusting orientation and speed to avoid collisions. The vehicle's path transitions from cyan to magenta.}
%%%	  \label{fig:fig4}
%%% \end{minipage}%
%%%% \hspace{0.5cm} % Add horizontal space between the two minipage
%%%\hspace{0.04\textwidth}% 
%%% \begin{minipage}[b]{0.48\textwidth}
%%%   		\centering
%%%		\includegraphics[width=0.6\columnwidth]{figs/gui_out2.png}
%%%
%%%  		\includegraphics[width=1.0\columnwidth]{figs/gui_out1a.png}
%%%    \caption{Parallel park-in maneuver. The vehicle's path transitions from cyan (initial position) to magenta (final position). Here, the vehicle must avoid going off the road when moving forward, before it backs into a slot between two cars. 
%%%} 
%%%  		\label{fig:fig5}
%%% \end{minipage}
%%%\end{figure*}

\begin{figure}[t!]
   		\centering
		\includegraphics[width=0.62\columnwidth]{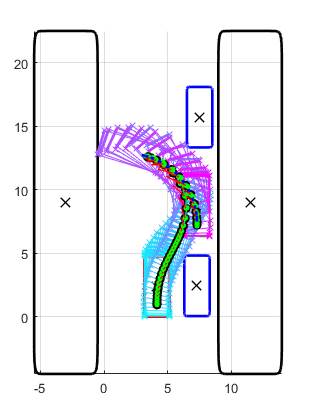}

  		\includegraphics[width=0.95\columnwidth]{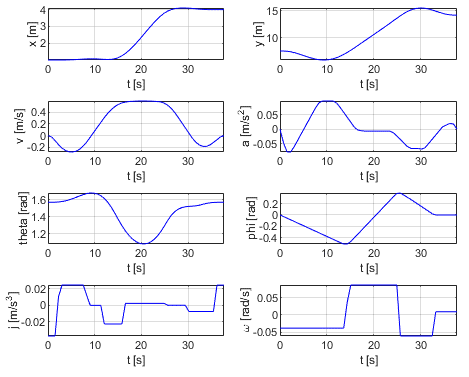}
    \caption{Parallel park-in maneuver. 
The host vehicle avoids going off road when moving forward, before backing into a slot between two cars.     
    The vehicle boundary transitions from cyan (initial position) to magenta (final position). 
} 
  		\label{fig:fig5}
% \end{minipage}
\end{figure}

%%\begin{figure}[t]
%%  \centering
%%  \includegraphics[width=0.5\textwidth]{figs/gui_out1.png}
%%%  \caption{caption}
%%%  \label{fig:fig1}
%%%\end{figure}
%%%
%%%\begin{figure}[t]
%%%  \centering
%%
%%  \includegraphics[width=0.5\textwidth]{figs/gui_out1a.png}
%%  \caption{caption}
%%  \label{fig:fig3}
%%\end{figure}
%%
%%
%%\begin{figure}[t]
%%  \centering
%%  \includegraphics[width=0.5\textwidth]{figs/gui_out2.png}
%%%  \caption{caption}
%%%  \label{fig:fig1}
%%%\end{figure}
%%%
%%%
%%%\begin{figure}[t]
%%%  \centering
%%
%%  \includegraphics[width=0.5\textwidth]{figs/gui_out1a.png}
%%  \caption{caption}
%%  \label{fig:fig4}
%%\end{figure}

In this section, results are presented for our optimal trajectory planning method 
in various parking scenarios. 
%
%In our implementation, the discretization of equation~\eqref{eq:eq:Ak_Bk_Ek_rk_int} was performed using SCVX toolbox [ref] in Matlab. A custom implementation of 
%equation~\ref{eq:min_traj} was written in Python using CVXPYGEN [ref], 
%a solver-generator that outputs efficient to complied C++. 
%
%In our implementation, 
To achieve this, a custom implementation of equation~\eqref{eq:min_traj} was written using CVXPYGEN~\cite{schaller2022embedded}, a Python-based solver-generator that produces efficient C++ output for minimizing optimization problems.
The discretization of equation~\eqref{eq:Ak_Bk_Ek_rk_int} was performed using the SCVX Toolbox~\cite{reynolds2020real} in Matlab, which was used to call the compiled solver. 
A custom graphical user interface (GUI) was created to specify vehicle and obstacle geometry, along with the kinematic constraints of equations~\eqref{eq:min_traj_d} to \eqref{eq:min_traj_f}. % in Matlab. 

Results are demonstrated in 
Figures~\ref{fig:fig2} to \ref{fig:fig5}, which illustrate the vehicle's path and the corresponding state and control variables over time. 
In each figure, the vehicle's path is depicted with a marker representing each waypoint, 
and a corresponding rectangle to indicate the vehicle's position and orientation at each step. 
Boundaries that define the obstacles are also shown.
The plotted variables include:
$x(t)$, the x-coordinate of the vehicle's position; 
$y(t)$, the  y-coordinate of the vehicle's position;
$v(t)$, the  vehicle's velocity;
$a(t)$, the  vehicle's acceleration;
$\theta(t)$, the  vehicle's orientation angle;
$\phi(t)$, the  steering angle;
$j(t)$, the jerk (rate of change of acceleration); and $\omega(t)$, the steering rate.

%\paragraph{Rear-In Perpendicular Parking Maneuver}

%\begin{figure}[H]
%    \centering
%    \includegraphics[width=0.8\textwidth]{fig2.png}
%    \caption{Rear-in perpendicular parking maneuver. The vehicle's path is shown with each waypoint represented as a rectangle, transitioning from cyan (initial position) to magenta (final position). The plots below show the state and control variables over time.}
%\end{figure}

Figure~\ref{fig:fig2} shows the trajectory generated for a rear-in perpendicular parking maneuver. 
Here, the \enquote{L}-shaped driving region is formed by four rectangular obstacles. 
%The plots below the path diagram illustrate the evolution of the vehicle's state and control variables over time. 
Plots of
the
state and control variable
 demonstrate how the vehicle adjusts its speed and steering to achieve a smooth and collision-free trajectory into the parking slot.
%
%\paragraph{Rear-In Parallel Parking Maneuver}
%
%\begin{figure}[H]
%    \centering
%    \includegraphics[width=0.8\textwidth]{fig3.png}
%    \caption{Rear-in parallel parking maneuver. The vehicle's path is shown with each waypoint represented as a rectangle, transitioning from cyan (initial position) to magenta (final position). The plots below show the state and control variables over time.}
%\end{figure}
%
A rear-in parallel parking maneuver is shown in 
Figure~\ref{fig:fig3}. % presents the results for .
In Figure~\ref{fig:fig5},  the results for a parallel park-in maneuver are shown. 
Here, the vehicle must avoid going off the road when moving forward, then it backs into a slot between two cars. 
%Figure~\ref{fig:fig5} also illustrates why a polygonal vehicle boundary was needed, 
%as gaps in the rectangular spacing are visible near the upper obstacle. 

%The use of a polygonal vehicle boundary prevents gaps that would occur with a rectangular boundary, ensuring a collision-free trajectory. 

%The vehicle's path is depicted with waypoints, and the obstacles are shown. The state and control variables are plotted, demonstrating how the vehicle maneuvers into the parallel parking slot. 
%The variables include x(t), y(t), v(t), a(t), $\theta$(t), $\phi$(t), j(t), and $\omega$(t).

%\subsection{Discussion}

These results demonstrate the effectiveness of our proposed method for generating optimal and collision-free trajectories. % for various parking scenarios. 
%The vehicle %successfully 
%navigates
%The approach provided 
Successful navigation
 into and out of parking slots was achieved while adhering to the constraints on state and control variables.
%The smoothness of the trajectories and the avoidance of obstacles validate the efficiency and reliability of the sequential convex optimization approach.

%\clearpage

\section{Conclusion}

This paper presented a novel method for optimal trajectory planning. 
% with collision avoidance for autonomous vehicle maneuvering. 
 Using sequential convex optimization, we generated efficient and collision-free trajectories for various parking scenarios. The proposed approach %effectively 
handles vehicle kinematics, obstacle avoidance, and sensor blind-spots, ensuring smooth and feasible paths. The results demonstrated the method's capability to navigate complex parking maneuvers, validating its practical applicability and efficiency. Future work may explore real-time implementation and integration with advanced sensor technologies to further enhance the system's robustness and performance.

\label{sec:concl}

%
%\bibliographystyle{abbrv}
% \bibliography{objects_references}  % need to put bibtex references in references.bib 
\bibliographystyle{ieeetran}
\bibliography{objects_references}  % need to put bibtex references in references.bib 

\end{document}